\documentclass[epjCONF]{svjour}
\usepackage{graphics}
\usepackage[varg]{txfonts} 
\usepackage[latin1]{inputenc}
\session-title{FUSION11}
\begin{document}
\title{Upper Limit on the molecular resonance strengths in the ${}^{12}$C+${}^{12}$C fusion reaction}
\author{X. Tang\inst{1}\fnmsep\thanks{\email{xtang1@nd.edu}} \and X. Fang\inst{1} 
\and B. Bucher\inst{1} \and H. Esbensen\inst{2} 
\and C. L. Jiang\inst{2} \and K. E. Rehm\inst{2} \and C. J. Lin\inst{3} }
\institute{Department of Physics and Joint Institute of Nuclear Astrophysics, University of Notre Dame, Notre Dame, IN 46556, USA \and Physics Division, Argonne National Lab, Argonne, IL 60439, USA \and Department of Nuclear Physics, China Institute of Atomic Energy, Beijing 102413, China}
\abstract{
Carbon burning is a crucial process for a number of important astrophysical scenarios.
The lowest measured energy is around E$_{\rm c.m.}$=2.1 MeV, 
only partially overlapping with the energy range of astrophysical interest. 
The currently adopted reaction rates are based on an extrapolation which is
highly uncertain because of potential resonances existing in the unmeasured 
energy range and the complication of the effective nuclear potential. 
By comparing the cross sections of the three carbon isotope fusion reactions, 
${}^{12}$C+${}^{12}$C, ${}^{12}$C+${}^{13}$C and ${}^{13}$C+${}^{13}$C, 
we have established an upper limit on the molecular resonance strengths 
in ${}^{12}$C+${}^{12}$C fusion reaction. 
The preliminary results are presented and 
the impact on nuclear astrophysics is discussed. 
} 
\maketitle
\section{Introduction}
\label{intro}
In 1960 Almqvist, Kuehner and Bromley discovered several resonances 
in collisions between ${}^{12}$C nuclei 
that provided the first evidence for the existence of nuclear molecules. 
For at least three energies, E$_{\rm c.m.}$=5.68, 6.00 and 6.32 MeV,
they observed resonances in the yield of collisional by-products:
protons, gamma radiation, alpha particles and neutrons. 
The resonances had characteristic widths of about 120 keV, which 
corresponds to a lifetime of 5$\times$10$^{-21}$ seconds,
a value more than 10 times greater than the typical time span 
for a simple collision\cite{ABK,bromley78}.
In the following years, as the measurements were pushed down towards lower energies, 
the discoveries of such resonances continued as far as the lowest measured energy. 
For instance, the most recent published measurement of 
the ${}^{12}$C+${}^{12}$C fusion reported a strong resonance at E$_{\rm c.m.}$=2.14 MeV. 
However, the reported width (less than 12 keV) is much less than the 
common 100 keV observed at higher energies\cite{sp07}.
\\
Carbon burning driven by the ${}^{12}$C+${}^{12}$C fusion, 
is a crucial process for a number of important astrophysical scenarios, 
such as the formation of white dwarfs, nucleosynthesis in massive stars, 
carbon ignition in type Ia supernovae and in superbursts\cite{gas07}. 
The important energy range spans from 1 MeV to 3 MeV in the center of mass frame, 
which is only partially covered by the measurements 
of the ${}^{12}$C+${}^{12}$C fusion reaction. The extrapolation is the only resource 
available to obtain the fusion reaction rate for the astrophysical applications. 
The current adopted ${}^{12}$C+${}^{12}$C fusion reaction rate is established based on
the modified S factor(S*), which is defined as,
\begin{equation}\label{nfdef}
S^{*}(E)=\sigma(E) E e^{\frac{87.21}{\sqrt{E}} + 0.46 E}.
\end{equation}
An averaged S* factor of 3$\times$10$^{16}$ MeV*b was obtained by fitting the data 
measured by Patterson\cite{patterson69}, Spinka\cite{spinka74} and Becker\cite{becker85}. 
This averaged value was extrapolated down to
the energies of astrophysical interest with a simple assumption 
that the averaged modified S factor is constant at the sub-barrier energies\cite{cf88}. 
\\
The strong and relatively narrow resonances continue down to the lowest 
energies for which measurements have been made. 
To present, there is nothing known about 
the exact energies and strengths of these resonances in the energy region 
which is significant in the astrophysical context. Therefore, 
the currently adopted carbon fusion rate based 
on the averaged S* factor is highly uncertain.
\\
Superbursts are long, energetic, rare explosions in Low Mass X-Ray Bursts. These bursts
are considered to be triggered by the unstable carbon burning in the 
ash left over from the rp-process on the surface of a neutron star. 
The ash is heated up with the heat
generated in the crust of a neutron star and ignites the carbon in the ash.
However, with the currently adopted carbon fusion rate, 
superbust models fail to explain the ignition 
of carbon at the column depth inferred from observation\cite{ke08}. 
Another difficulty is that the amount of ${}^{12}$C
in the ash does not seem to be sufficient to trigger the superburst.
Inspired by the strong resonance observed at E$_{\rm c.m.}$=2.14 MeV, 
a hypothetical resonance with a similar strength was proposed to enhance
the fusion reaction rate at T$_9$= 0.5 K with a factor of over two orders of 
magnitude 
and thereby alleviate the discrepancy
between superburst models and observations\cite{cooper09}.

\section{Empirical relationship among the carbon isotope fusion reactions}
\label{cifr}
\begin{figure}
\resizebox{1.0\columnwidth}{!}{%
\includegraphics{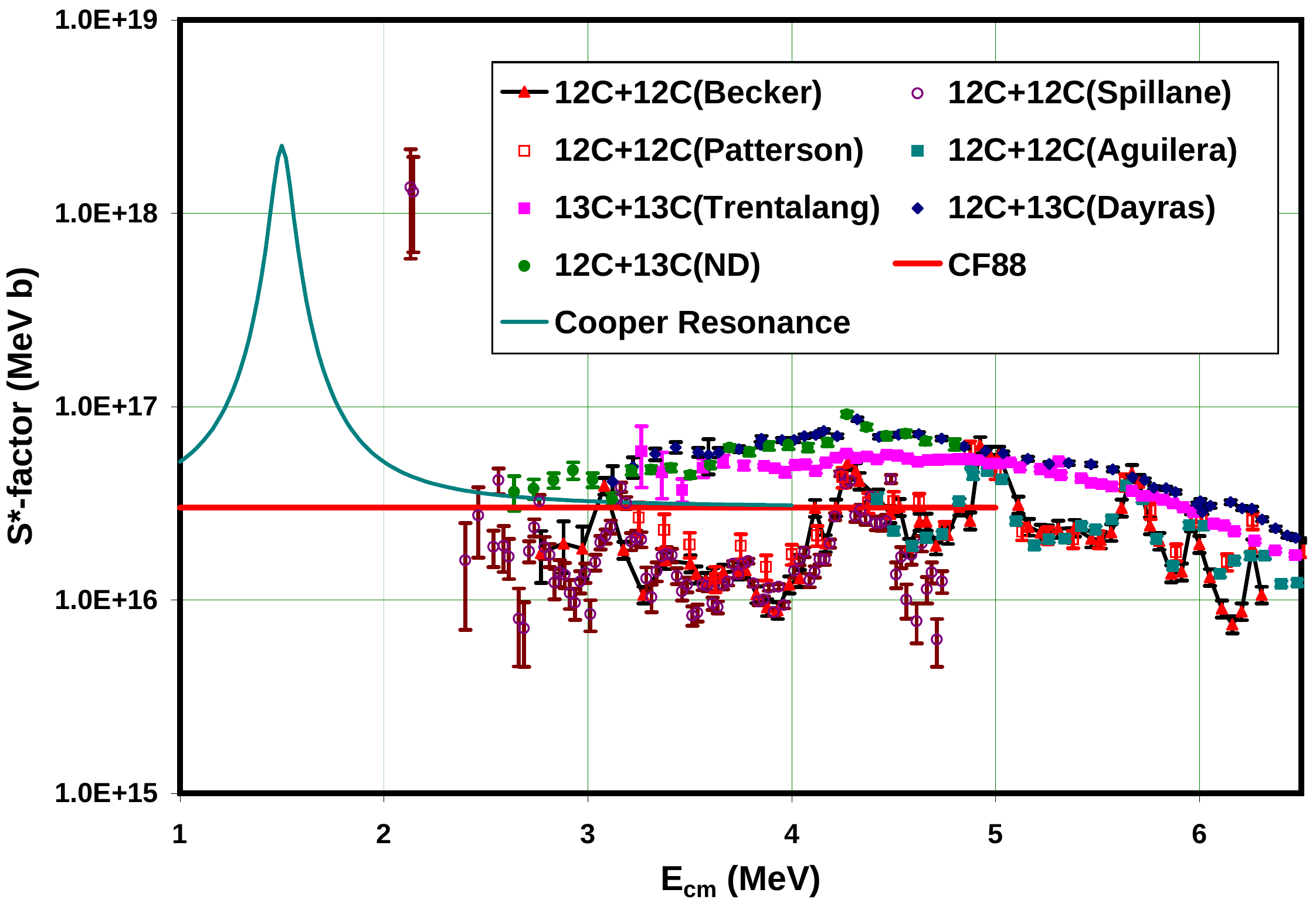}}
\caption{The experimental S* factors of three carbon isotope fusion reactions, 
${}^{12}$C+${}^{12}$C\cite{sp07,patterson69,becker85,agui06}, 
${}^{12}$C+${}^{13}$C\cite{dayras75,masa09} and ${}^{13}$C+${}^{13}$C\cite{trentalange88}. 
The averaged S* factor of ${}^{12}$C+${}^{12}$C(S*=3$\times$10$^{16}$ MeVb), shown as red line, 
is recommended based on 
the data of Becker, Patterson and Spinka (not shown here). 
There are some data with large error bars in Spillane's data. 
In this paper, only those with relative errors less than 70\% are shown.
The hypothetical resonance proposed by Copper {\it et al.} is also shown.}
\label{fig:1}       
\end{figure}

In contrast to the striking resonances in the ${}^{12}$C+${}^{12}$C fusion reaction, 
other carbon isotope fusion reactions, such as ${}^{12}$C+${}^{13}$C and ${}^{13}$C+${}^{13}$C, 
behave more regularly. Little resonance feature has been observed in 
these two systems~\cite{dayras75,trentalange88}. The shape of the fusion cross sections 
at sub-barrier energies is primarily dominated by the Coulomb barrier penetration effect. 
To remove this effect and reveal more details of nuclear interaction, 
the cross sections of all three carbon fusion systems are 
converted into the S* factors using eq.\ref{nfdef}. The advantage of 
using this conversion over the traditional astrophysical S factor is 
that the cross section ratios among the three systems are preserved in this approach.
\\
The experimental S* factor for all three carbon isotope fusion systems are 
shown in Fig.\ref{fig:1}. There are several important 
features in these carbon isotope fusion systems:
\begin{enumerate}
\item 
The ${}^{12}$C+${}^{12}$C cross sections are bounded from above by 
the cross sections of the other two carbon isotope fusion systems.
\item All the maxima of the ${}^{12}$C+${}^{12}$C fusion cross section 
(E$_r$=3.08, 4.28, 4.92, 5.67, 5.96 and 6.26 MeV) 
match remarkably well with the relatively smooth cross sections observed for the other two isotope combinations, especially 
${}^{13}$C+${}^{13}$C.
\item Overall, the ${}^{12}$C+${}^{13}$C cross sections are the highest among the three carbon isotope
fusion systems. The deviation between ${}^{12}$C+${}^{13}$C and ${}^{13}$C+${}^{13}$C 
occurs in the energy of 3.5 to 5 MeV 
(excluding the 4.25 MeV resonance in ${}^{12}$C+${}^{13}$C) and is less than 30\%. 
\item The 4.25 MeV resonance in ${}^{12}$C+${}^{12}$C is 
also present in ${}^{12}$C+${}^{13}$C.
\end{enumerate}

The Nogami-Imanishi model has suggested the coupling effect to the ${}^{12}$C(2+,4.44 MeV)
is crucial to the formation of ${}^{12}$C+${}^{12}$C molecular resonances\cite{imanishi68}.
In the  ${}^{12}$C+${}^{13}$C and ${}^{13}$C+${}^{13}$C fusion reactions, 
the valence neutron of ${}^{13}$C introduces an additional freedom for 
excitation besides the ${}^{12}$C core excitation. 
Despite this complication, 
the maximum deviation among the ${}^{12}$C+${}^{12}$C peak cross sections, 
${}^{12}$C+${}^{13}$C and ${}^{13}$C are less than 30\%. 
Based on this strong correlation, we claim:
\begin{enumerate}
\item
All major resonances have the same reaction mechanism;
\item
This reaction mechanism should be very similar to the process
which fuses ${}^{12}$C+${}^{13}$C and ${}^{13}$C+${}^{13}$C.
\item
The effect arising from the valence neutron in ${}^{13}$C is not
strong compared to this reaction mechanism. 
\end{enumerate}
The Imanishi calculation may provide a qualitative explanation to the 
${}^{12}$C+${}^{12}$C experimental data. However, no
quantitative description  has been achieved yet.  

\section{The ${}^{12}$C+$^{13}$C experiment at Notre Dame and test of extrapolating models 
for heavy ion fusion reactions at extreme sub-barrier energies}
\begin{figure}
\resizebox{1.0\columnwidth}{!}{%
\includegraphics{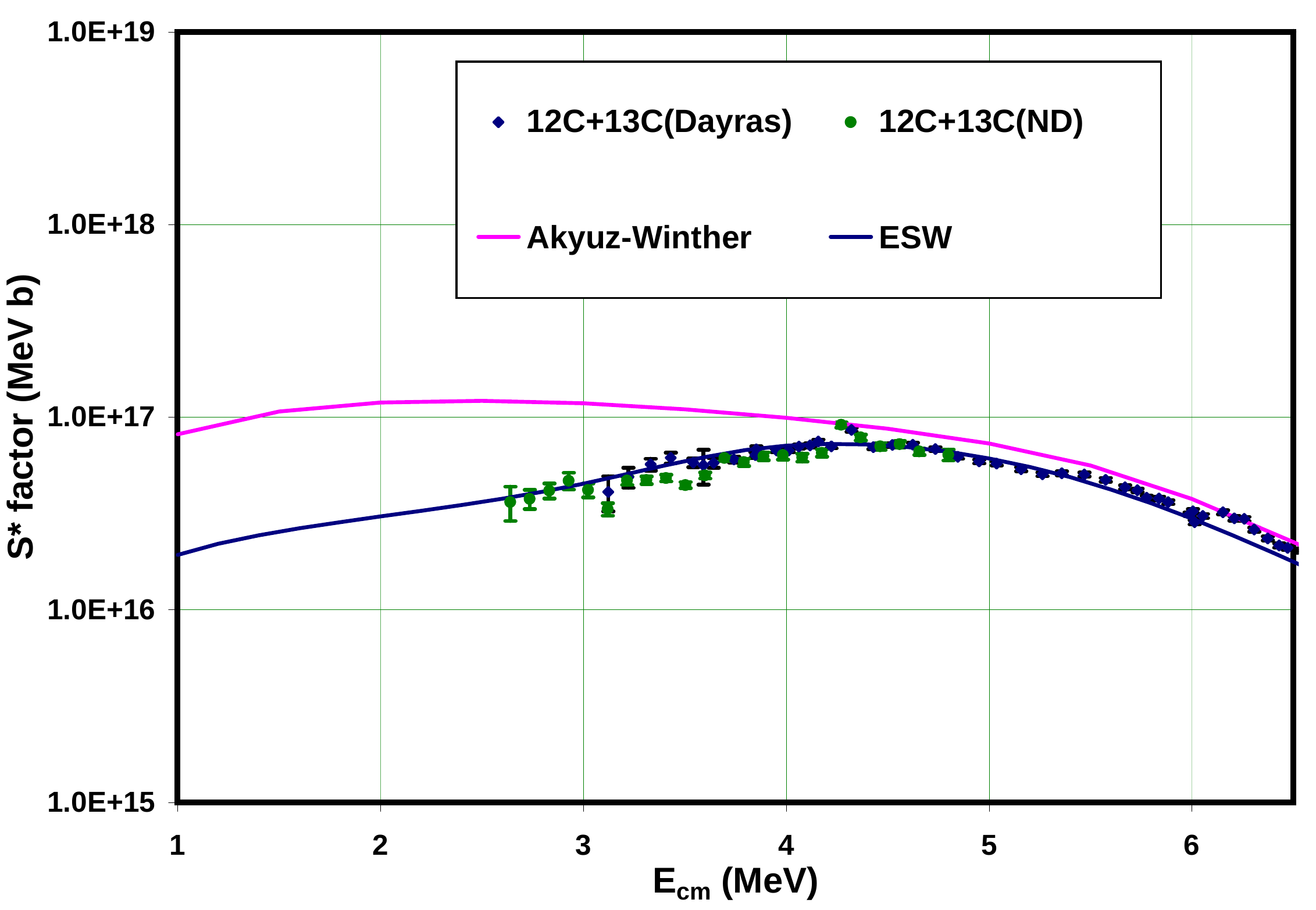}}
\caption{The experimental S* factors for ${}^{12}$C+${}^{13}$C\cite{dayras75,masa09}. 
The ESW fit to the Dayras data (blue points)
is shown as blue line while the Aky$\ddot{\rm u}$z-Winther calculation is shown as magenta line.
The ESW prediction match the new measurement studied at Notre Dame (green points).}
\label{fig:2}       
\end{figure}

Up to now, measurements of ${}^{12}$C+${}^{12}$C have reached down to 2.1 MeV at which 
the cross section is below 1 nb. However, measurements of 
${}^{12}$C+${}^{13}$C and ${}^{13}$C+${}^{13}$C stop at 
energies above 3 MeV. The lowest measured cross sections are at the level of $\mu$b. 
It would be interesting to extend the measurements of ${}^{12}$C+${}^{13}$C 
and ${}^{13}$C+${}^{13}$C towards lower energies 
and  examine the empirical correlations among the three systems.
\\
The correlations among the three carbon isotope systems provide a possibility to establish 
an upper limit for the important ${}^{12}$C+${}^{12}$C. 
If all the major resonances in ${}^{12}$C+${}^{12}$C, including those to be discovered at lower energies, 
are driven by the same reaction mechanism, one can expect to establish an extrapolation 
based on the resonant cross sections observed at higher energies. To achieve a more reliable 
extrapolation within the energy range of 1 to 3 MeV, it is necessary to push the measurements 
of the carbon isotope fusion reactions towards lower energies.
\\
As the first step, we have studied the ${}^{12}$C+${}^{13}$C fusion within the center 
of mass energy range of 2.6 MeV to 4.8 MeV using the FN tandem accelerator at Notre Dame. 
In this experiment, 
the ${}^{12}$C(${}^{13}$C,${}^{24}$Na)p channel was studied via decay spectroscopy. 
The total fusion cross section was obtained by correcting the measured cross section using 
the branching ratio from Ref.\cite{dayras75}. 
The lowest cross section reached in this measurement is about 20 nb, a factor of 50
smaller than which Dayras achieved\cite{dayras75}. 
The experimental detail can be found in Ref.\cite{masa09}. 
The new ${}^{12}$C+${}^{13}$C data are shown 
together with the Dayras measurement in fig. \ref{fig:1} and \ref{fig:2}. 
In general, the shape of the new data agrees with the Dayras measurement 
except in the range of 3.3 to 3.5 MeV 
where the new data is a little lower than the Dayras measurement. It may be explained as 
a fluctuation in the branching ratio incurred by the nuclear structure in the compound nucleus,
${}^{25}$Mg. It is also observed in fig. \ref{fig:1} that 
the new ${}^{12}$C+${}^{13}$C data still lie above the ${}^{12}$C+${}^{12}$C 
cross sections measured by Spillane {\it et al.} within the overlapped energy range. 


The fusion systems with smooth cross sections, including  ${}^{12}$C+${}^{13}$C 
and ${}^{13}$C+${}^{13}$C, can be reasonably described by
coupled-channel calculations which include the couplings between 
the relative motion of the colliding nuclei and several nuclear intrinsic 
motions as well as transfer reactions. One challenge in this approach
is the determination of the effective nuclear potential. 
So far, there is no universal potential which may provide reliable extrapolation
down to the extreme sub-barrier energies. 
For example, the coupled channel calculation with the Aky$\ddot{\rm u}$z-Winther potential,
a global potential with a Woods-Saxon shape, provides a good description of the fusion cross 
sections around the Coulomb barrier. However, it was found in the hindrance study that, at lower energies,
the calculation  always over-predicts the fusion cross sections\cite{misicu07}. 
Right now, there is a ongoing effort trying to model all three carbon isotope fusion reactions using 
a unified double folding potential\cite{henning}. 
\\
In contrast to the standard approach, 
the Equivalent Square Well (ESW) model has been chosen traditionally 
to model the sub-barrier fusion cross sections and provide extrapolation 
within the energy ranges of astrophysical interest. 
Although the physical justification of the model may 
not be as clear as the coupled channel calculations used in the sub-barrier fusion study, 
this model has achieved fits to measured data that are 
superior to those obtained from most other models\cite{barnes85}.
To examine the predictive power of the ESW model for the carbon isotope fusion reactions, 
the ESW extrapolation determined in an earlier measurement 
at higher energies\cite{dayras75} is shown in Fig. \ref{fig:2} 
together with the new ${}^{12}$C+${}^{13}$C data.
The excellent agreement proves the superior predictive power of the ESW model.
As a comparison,  a couple channel calculation using the Aky$\ddot{\rm u}$z-Winther potential was carried
out with the code, CCFULL\cite{hagino99}.  The ${}^{13}$C(3/2$^-$, 3.684 MeV) 
with strong B(E2) is included in the calculation. The result is shown in Fig. \ref{fig:2}.
The calculation significantly over predicts the fusion cross section at the energies less than 4 MeV.
For example, the fusion cross section at 2.6 MeV is over predicted by a factor of 3.4. 

\section{Upper limit on the ${}^{12}$C+${}^{12}$C molecular resonance strengths}
\begin{figure}
\resizebox{1.0\columnwidth}{!}{%
\includegraphics{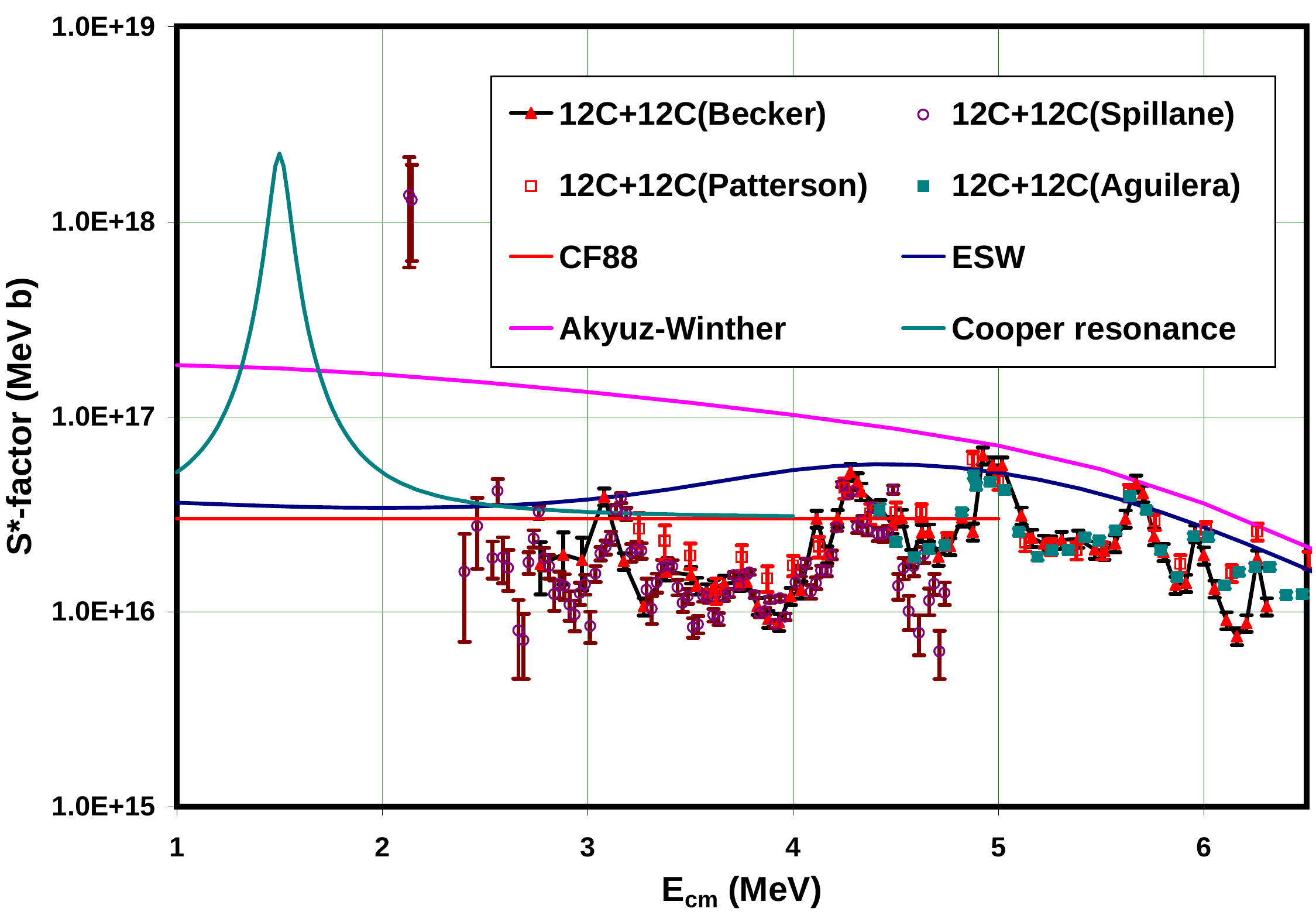}}
\caption{The experimental S* factors for ${}^{12}$C+${}^{12}$C\cite{sp07,patterson69,becker85,agui06}. 
The ESW model is fit to the major peak cross sections at  
E$_r$=3.08, 4.28, 4.92, 5.67, 5.96 and 6.26 MeV. 
The ESW fit is shown as blue line.
The Aky$\ddot{\rm u}$z-Winther calculation is also shown as magenta line. 
It is clear that the resonant cross section at 1.5 and 2.14 MeV 
are too high to be realistic.
}
\label{fig:3}       
\end{figure}

The resonance strength is directly related to the peak cross section
since all the nuclear molecular states are expected to have a similar width. 
The correlation between the ${}^{12}$C+${}^{12}$C peak cross sections and 
the cross sections of the other two carbon isotope fusion systems 
provides a possibility to model the resonance strengths.
The ESW model has been applied to fit those major peak cross sections 
( E$_{\rm r}$=3.08, 4.28, 4.92, 5.67, 5.96 and 6.26 MeV) which match 
the ${}^{12}$C+${}^{13}$C and ${}^{13}$C+${}^{13}$C
fusion cross sections. The ESW extrapolation is shown in Fig. \ref{fig:3} together with 
the experimental data. 
Even though the fit is based on the peak cross sections above 3 MeV, the 
extrapolation to lower energies provide an 
upper bound on most of Spillane's data well, except the strong resonance observed at 2.14 MeV.
The resonant cross section at this energy is about a factor of 40 higher than the ESW extrapolation.
The preliminary result from the ${}^{12}$C+${}^{12}$C 
experiment\cite{jim10} at Naples supports our finding. 
\\
Another, even safer, upper limit can be estimated by coupled-channels calculation that 
are based on the he Aky$\ddot{\rm u}$z-Winther potential. 
This approach is known to 
over-predict the fusion cross sections at extreme sub-barrier energies.  
The calculation is carried out using CCULL\cite{hagino99} by including 
the coupling to the  ${}^{12}$C(2$^+$,4.44 MeV).
The result is also shown in Fig. \ref{fig:3}. 
The two resonances at 1.5 MeV 
and 2.14 MeV are about a factor of 10 higher than 
the Aky$\ddot{\rm u}$z-Winther calculation.

We cannot rule out the possibility of a strong resonance 
in the unmeasured energy range, formed by a mechanism different 
than the ones we have discussed here. The established upper limit
cannot account for such a resonance.  
However, there is no such evidence thus far.
In principle, there are some sharp resonances existing 
in the ${}^{24}$Mg which decay by gamma emission. 
It is possible that those resonant  cross sections 
are much higher than the ESW extrapolation, but the
resonance strengths are too small to affect 
the ${}^{12}$C+${}^{12}$C fusion reaction. 

\section{Impact to nuclear astrophysics}
Even though we do not know the exact location of the potential resonance in
the unexplored energy range, with the established upper limit,
we can claim the real ${}^{12}$C+${}^{12}$C fusion rate is much less than 
Cooper {\it et al.} proposed. Taking the over-predicted S* factors 
calculated with the Aky$\ddot{\rm u}$z-Winther potential, the real rate should not be more than 
a factor of 6 stronger than the currently adopted fusion rate 
based on the S* factor in the energy range of 1 to 3 MeV. 
The upper limit established with the ESW fit, the best fit we have achieved, 
is only 20\% higher than the currently adopted rate. The  
averaged S* factor of Spillane's data tends to suggest a smaller average 
than the current value. Therefore, the more than a factor of 40 enhancement as 
proposed in Ref.\cite{cooper09} is impossible. 
\\
The upper limit posts an additional constraint on the current superburst models.
Since the carbon burning rate can not be as high as Cooper {\it et al.} proposed, there must 
be some unknown physics processes not being included in the superburst models.
For example, an additional heating process in the neutron star crust 
could raise the peak temperature in the ash to meet the required ignition condition.
It is also possible that some process other than carbon burning triggers 
the superburst.

\section{Summary and outlook}
Carbon fusion is an important process.
The molecular resonance complicates the prediction of the carbon fusion
rate at the energies of astrophysical relevance. 
By comparing the cross sections of three carbon isotope fusion reactions, 
${}^{12}$C+${}^{12}$C, ${}^{12}$C+${}^{13}$C and ${}^{13}$C+${}^{13}$C, 
an empirical relationship among the three systems is found. With this 
relationship and the Equivalent Square Well (ESW) model, 
for the first time, the resonant cross sections in 
the ${}^{12}$C+${}^{12}$C fusion reactions is quantitatively described.
Consequently, an upper limit on the molecular resonance strengths 
in ${}^{12}$C+${}^{12}$C fusion reaction is established.
This upper limit serves 
an additional constraint for the superburst models.
\\
To provide a more accurate prediction at extreme sub-barrier energies, it is 
urgent to improve the experimental technique so that we can 
push the measurements of the carbon isotope 
fusion reactions towards lower energies. 
Right now, the Notre Dame
group is building a 5 MV single end accelerator with an ECR source to
provide more than 40 p$\mu$A carbon beam. We are also improving detection 
techniques to increase efficiency as well as selectivity \cite{jiang11}. 
These approaches are expected to provide better 
experimental data in the near future. 
The proposed underground laboratory with 
a high current heavy ion accelerator will be the ideal place to
provide the ultimate answer to the carbon fusion rate. 

\section{Acknowledgment}
This work is supported by the NSF under Grant No. PHY-0758100 and PHY-0822648,
the DOE office of Science through DE-AC02-06CH11357, 
the National Natural Science Foundation of China under Grant No. 11021504, 
and the University of Notre Dame. 
X.T. thanks Prof. Wiescher, Prof. Kolata and Dr. Beard at Notre Dame, 
Prof. Tostevin at Surrey, Prof. Trentalange at UCLA
and Prof. Brown at MSU for the inspiring discussions during the work. 
X.T. also thanks Prof. Hagino for his kind helps 
on using his code, CCFULL.

\end{document}